\documentclass[aps,prl,preprint,groupedaddress]{revtex4}
\usepackage{amsmath}
\usepackage{graphicx}

\begin{document}

\title{Thermal rectification in thickness-asymmetric graphene nanoribbons}
\author{Wei-Rong Zhong$^{1}$}
\email{wrzhong@hotmail.com}
\author{Wei-Hao Huang$^{1}$}
\author{Xi-Rong Deng$^{1}$}
\author{Bao-Quan Ai$^{2}$}
\email{aibq@scnu.edu.cn}
\affiliation{$^{1}$\textit{Department of Physics and Siyuan Laboratory, College of
Science and Engineering, Jinan University, Guangzhou 510632, P. R. China}\\
$^{2}$\textit{Laboratory of Quantum Information Technology, ICMP and SPTE,
South China Normal University, Guangzhou, 510006 P. R. China}}
\date{\today }

\begin{abstract}
Thermal rectification in thickness asymmetric graphene nanoribbons
connecting single-layer with multi-layer graphene is investigated by using
classical nonequilibrium molecular dynamics. It is reported that the
graphene nanoribbons with thickness-asymmetry have a good thermal
rectification. The thermal rectification factor depends on temperature as
well as the thickness-ratio of the two-segment. Our results provide a direct
evidence that the thermal rectifier can be achieved in a nanostructure
crossing two- and three-dimension.
\end{abstract}

\maketitle


In the past decade, the studies of the mechanism of heat conduction lead to
potentially interesting applications based on the possibility to control the
heat flow. Thermal rectification has potential applications in nanoscale
thermal management such as on-chip cooling and energy conversion by
controlling the heat transport. It's also fundamental in several recently
proposed the schemes of \textquotedblleft thermal
circuits\textquotedblright\ or information processing using phonons \cite%
{baowen1, baowen2, baowen3}. Since the thermal rectifier based on Morse
potential has been proposed \cite{peyrard}, many thermal rectifiers have
been revealved\ in various structures. For example, the thermal diode by
coupling two nonlinear lattices \cite{baowen1, dhe}, the asymmetric
nanotubes and graphene \cite{baowen2, wug, huj1, alaghemandi}, anharmonic
graded mass crystals \cite{pereira, yangn2} and a spin-boson nanojunction
model \cite{segal}. Inspired by these theoretical studies, Chang and
co-workers have produced a microscopic solid-state thermal rectifier based
on carbon nanotubes \cite{chang}. Correspondingly, for the study of thermal
rectifier, an immediate goal is to find a structure with good thermal
rectification factor.

Up to now, although the rectification factor in asymmetric nonlinear
Frenkel-Kontorova(FK) lattices has been reported to be about 100 \cite%
{baowen2}, the rectification observed in experiment is only 0.03 $\sim $
0.07 \cite{chang}. The rectifying coefficient of oxide thermal rectifier is
also only 0.4 \cite{terasaki}, which are far smaller than the predicted
value. The theoretical rectification of asymmetric carbon nanotubes is about
0.1 $\sim $ 0.12 \cite{wug}. Recently, N. Yang et al. \cite{yangn} and J. Hu
et al. \cite{huj1} reported a significant rectification of 3.0\ and 1.2,
respectively, in asymmetric single-layer graphene nanoribbons(GNRs).
However, the asymmetric GNRs of triangular shapes only contains one or two
atoms in one end. Whether this kind of GNRs can be easily fixed on the
substrate is still not clear. From the view of experiment, a few-layer
graphene can be fabricated more easily than a single-layer graphene \cite%
{ghosh, geim, balandin}. A thermal diode based on the few-layer graphene
should be more popular and its the potential applications may be enormous.
Furthermore, a thermal rectification based on the few-layer GNRs can be more
helpful to understand the thermal mechanism from two- to three-dimensional
materials.

In this letter, we build an asymmetric two-segment GNRs, as shown in Fig.1,
which is consisting of a multi-layer graphene in one segment and a
single-layer graphene in the other segment. In our studies, the right
segment is fixed as a single-layer graphene and the thickness of the right
segment can be changed through adding the layers. Here we will display that
this asymmetric structure also performs a good rectification. Additionally,
the thickness-asymmetric GNRs provide more alternative options for
controlling the rectification factor through changing the number of layers
in the multi-layer segment. The obtained results are of significance for
understanding the asymmetry and thickness effect on the rectification
factor. Therefore, this may open up multi-layer graphene's applications in
the nanoscale thermal diode.

In our simulations, we have used the Tersoff-Brenner potential \cite{tersoff}
for the C-C bonding interactions in intra-plane and the Lennard-Jones
potential \cite{girifalco} for van der Waals interactions in the
inter-plane. The Nos\'{e}-Hoover thermostats are applied to the left and
right end of GNRs \cite{huj1, wug, zhong}. $T$ ($T_{L}$ or $T_{R}$), the set
temperatures of the heat baths, are placed in the two ends of the graphene.
The average temperature is denoted as $T_{0}=(T_{L}+T_{R})/2,$ and the
temperature difference is $\Delta T=(T_{L}-T_{R})/2$. In order to avoid the
spurious global rotation of the graphene in the simulations, as shown in
Fig.1, we use fixed boundary conditions in the two ends of the graphene (L
and R). The fixed region and the heat baths occupy one layer and four layers
of atoms, respectively. Free boundary conditions are applied to the other
boundaries.

\begin{figure}[htbp]
\begin{center}\includegraphics[width=8cm,height=6cm]{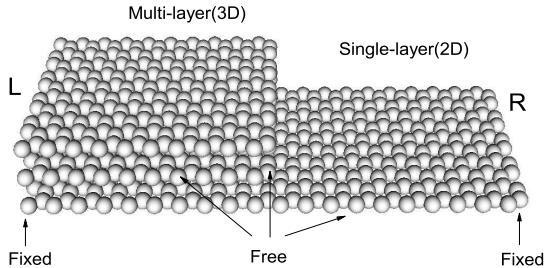}
  \end{center}
  \caption{Schematic diagram
showing the multi- and single-layer zigzag edge graphene coupling to two
heat baths. The two end of the graphene is fixed and the other boundaries
are free. The layers can be changed in multi-layer segment and the right
part is fixed as a single-layer. The total length of the GNRs is 5nm.}
   \label{}
\end{figure}

We integrate the equations of motion for atoms by the Verlet method \cite%
{rafii}. The total heat flux injected from the heat bath to the system can
be obtained by $J=\underset{i}{\sum }\left[ -\Gamma p_{i}^{2}/m_{i}\right]
=-3\Gamma Nk_{B}T(t),$where the subscript $i$ runs over all the particles in
the thermostat \cite{wug}. The direction of the flux $J_{L\rightarrow R}$
(from L to R) is positive and the direction of the flux $J_{R\rightarrow L}$
(from R to L) is negative. The thermal conductivity is calculated from the
well-known Fourier's law $\kappa =Jl^{2}/(\Delta TV)$, in which $l$ and $V$
are the length and the volume of the total GNRs, respectively. The thickness
of each layer of the graphene is 0.335nm. The thermal rectification factor
is defined as $R_{0}=|J_{L\rightarrow R}+J_{R\rightarrow
L}|/|J_{L\rightarrow R}|\times 100\%$. For comparing edge chirality
dependent on the thermal rectification, an armchair and a zigzag graphene
are investigated in parallel. The widths of the armchair and the zigzag
graphene are 1.6 and 2.1nm, respectively. The thickness asymmetry of the
GNRs is $\gamma =N_{L}/N_{R}$, where $N_{L}$ and $N_{R}$ are the numbers of
the layers at left segment and right segment, respectively.

As illustrated in Fig. 2, we present the heat current $J$ versus $\Delta T$
for both armchair and zigzag graphene. Here we use an asymmetric GNRs with a
two-layer graphene at the left segment and a single-layer graphene at the
right segment. When $\Delta T<0$, the heat current increases rapidly with $%
\Delta T$, while in the region $\Delta T>0,$ the heat current increases
slowly with $\Delta T$, so the system behaves as a thermal rectifier. The
inset in Fig. 2 shows that the rectification factor increases with the
temperature difference. The thermal rectification can also be described from
the thermal conductivity in different directions. We have investigated the
thermal conductivity of the thickness-asymmetric GNRs and found significant
thermal rectification. It is shown in Figure 3 that the thermal conductivity
from the multi-layer graphene (L) to the single-layer graphene (R) is less
than that from R to L. In calculating the thermal conductivity of
thickness-asymmetric GNRs, the thickness ($h$) is taken as the average
thickness of GNRs, which is given by $h=V/(wl)$, where $V,$ $l$ and $w$ are,
respectively, the volume, the length and the width of GNRs.

\begin{figure}[htbp]
\begin{center}\includegraphics[width=8cm,height=6cm]{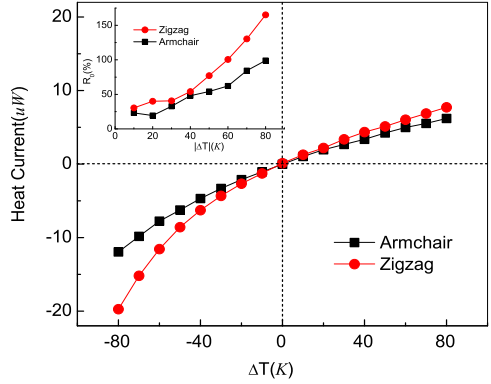}
  \end{center}
  \caption{Heat current as a function of the temperature
difference for armchair and zigzag edge asymmetric graphene nanoribbons. The
inset is the temperature difference dependence of the thermal rectification
factor. The average temperature $T_{0}=320K$.}
   \label{}
\end{figure}

\begin{figure}[htbp]
\begin{center}\includegraphics[width=8cm,height=6cm]{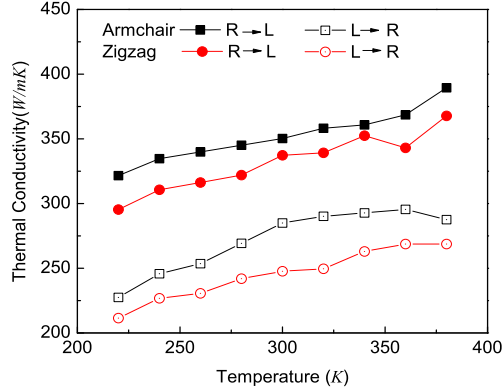}
  \end{center}
  \caption{Temperature
dependence of the thermal conductivity in GNRs for two opposite directions
(R to L and L to R). The temperature difference $\Delta T=20K$.}
   \label{}
\end{figure}

In Ref.\cite{wug}, the overlapping of the phonon spectra of the two atoms
near the connecting parts is used to explained the thermal rectification in
carbon nanotubes. It has been stated that in this real system, the
relationship between the overlap area and the absolute value of the heat
flux is the same as that in the one dimensional nonlinear lattice systems
\cite{baowen1, baowen2}, in which it is found that matching and/or
mismatching of the energy spectra near the interface is the underlying
mechanism of the rectification. In the final analysis, the rectification
origins from the asymmetric scattering of phonons through two opposite
directions. As shown in Fig. 1, when the phonons transfer from single-layer
(R side) to multi-layer (L side), i.e., from two- to three-dimensional
discontinuous structure, they have more channels to go through smoothly. On
the contrary, the phonons from multi-layer (L side) to single-layer (R
side), have to crush up near the junction, where three-dimension is
connected with two-dimension, and then the heat current is suppressed
significantly.

Figure 3 also shows that thickness-asymmetric GNRs have the rectifying
effect in a wide temperature range from 220k to 380k. The rectification
factor depends on the average temperature ($T_{0}$) as well as on the
thickness-ratio ($\gamma $).

As shown in Fig. 4, the rectification factor decreases with the average
temperature for different temperature differences. This can be explained as
the weaken effect of Umklapp scattering. The scattering of phonons increases
steadily with the temperature, and the asymmetry will be broken. When the
temperature increases to the value that Umklapp scattering in the
single-layer graphene approaches that in the junction, the rectification
factor goes to zero.
\begin{figure}[htbp]
\begin{center}\includegraphics[width=8cm,height=6cm]{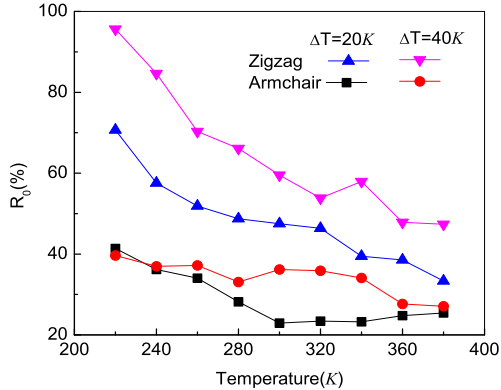}
  \end{center}
  \caption{Temperature
dependence of thermal rectification factor for two temperature difference. $%
T_{0}=320K$.}
   \label{}
\end{figure}

At small temperature difference $\Delta T=20K$, figure 5 shows that the
increasing of asymmetry can improve the thermal rectification factor. The
rectification effect is usually enhanced as the asymmetry of GNRs increases.
Interestingly, at large temperature difference $\Delta T=60K$, the
rectification factor decreases with the asymmetry of GNRs. Here we propose
an elementary explanation on this phenomenon. At large temperature
difference, when the number of the layers increases, the thickness strongly
suppresses the thermal conductivity of the GNRs \cite{zhong}. The heat
current cannot flow through the GNRs easily in the two directions, the
asymmetry dependence of GNRs on the current will become weaker, and then the
rectification will decrease. This result provides an important illumination
that the temperature difference is also a factor for improving thermal
rectification.
\begin{figure}[htbp]
\begin{center}\includegraphics[width=8cm,height=6cm]{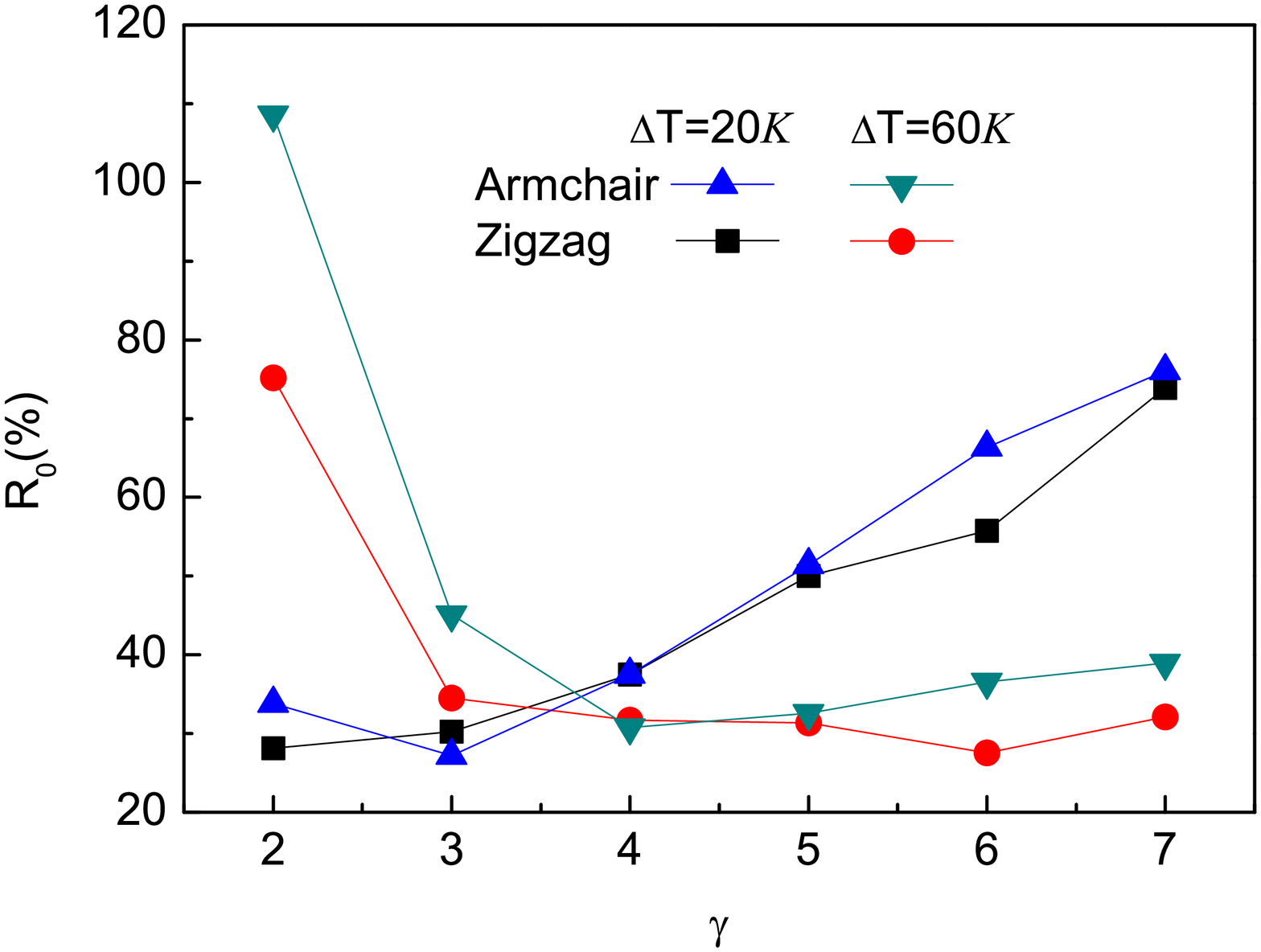}
  \end{center}
  \caption{Thickness ratio $%
\protect\gamma $ dependence of thermal rectification factor for two
temperature differences 20 and 60K, respectively. T$_{0}$=320K.}
   \label{}
\end{figure}

In summary, we have studied the thermal rectification of
thickness-asymmetric GNRs by using classical nonequilibrium molecular
dynamics. The calculated significant thermal rectification effect in
thickness-asymmetric GNRs is on the same order of asymmetric single-layer
graphene in Ref. \cite{huj1}. We have also demonstrated that armchair and
zigzag edges graphene perform similarly thermal rectification effect. The
rectification factor decreases with the temperature. The thickness-ratio
between the left and the right segment has a negative effect on the thermal
rectification factor at large temperature difference and a positive effect
at small temperature difference. Different from the previous thermal
rectifiers in graphene, we demonstrate another structure crossing two- and
three-dimension, which may be useful for clear understanding the thermal
transport in nanoscale materials from low- to high-dimension.

\begin{acknowledgments}
ZWR thanks Y. Xiao at City Univ. of Hong Kong for his contribution in
plotting Fig.1. This work was supported in part by the National Natural
Science Foundation of China (Grant Nos.11004082 and 11175067), the Natural
Science Foundation of Guangdong Province, China (Grant Nos.10451063201005249
and S201101000332) and the Fundamental Research Funds for the Central
Universities, JNU (Grant No.21609305).
\end{acknowledgments}


\begin{thebibliography}{99}
\bibitem{baowen1} B. Li, Lei Wang, and Giulio Casati, Appl. Phys. Lett.
\textbf{88}, 143501 (2006).

\bibitem{baowen2} B. Li, L. Wang, G. Casati, Phys. Rev. Lett. \textbf{93},
184301 (2004).

\bibitem{baowen3} L. Wang, B. Li, Phys. Rev. Lett., \textbf{99}, 177208
(2007); L. Wang, B. Li, Phys. World, \textbf{3}, 27 (2008).

\bibitem{peyrard} M. Terraneo, M. Peyrard and G. Casati, Phys. Rev. Lett.,
\textbf{88}, 094302 (2002).

\bibitem{dhe} B. Hu, D. He, L. Yang, and Y. Zhang, Phys. Rev. E, \textbf{74}%
, 060101 (2006); B. Hu, L. Yang, and Y. Zhang, Phys. Rev. Lett., \textbf{97}%
, 124302 (2006).

\bibitem{huj1} J. Hu, X. Ruan, and Y. P. Chen, Nano Lett., \textbf{9},
2730-2735 (2009).

\bibitem{wug} G. Wu and B. Li, Phys. Rev. B \textbf{76}, 085424 (2007); J.
Phys. Condens. Matter 20, 175211 (2008).

\bibitem{alaghemandi} Mohammad Alaghemandi, Frederic Leroy, Florian
Muller-Plathe, and Michael C. Bohm Phys. Rev. B, \textbf{81}, 125410 (2010).

\bibitem{yangn2} N. Yang, N. Li, L. Wang, and B. Li, Phys. Rev. B \textbf{76}%
, 020301(R) (2007).

\bibitem{pereira} E. Pereira, Phys. Rev. E \textbf{82,} 040101(R) (2010).

\bibitem{segal} D. Segal, A. Nitzan, Phys. Rev. Lett. \textbf{94}, 034301
(2005).

\bibitem{chang} C. W. Chang, D. Okawa, A. Majumdar, and A. Zell, Science
314, 1121 (2006).

\bibitem{terasaki} W. Kobayashi, Y. Teraoka, and I. Terasaki, Appl. Phys.
Lett. \textbf{95}, 171905 (2009); D. Sawaki, W. Kobayashi, Y. Moritomo, and
I. Terasaki, Appl. Phys. Lett. \textbf{98}, 081915 (2011).

\bibitem{yangn} N. Yang, G. Zhang, and B. Li, Appl. Phys. Lett. \textbf{95},
033107 (2009).

\bibitem{geim} R. R. Nair, P. Blake, A. N. Grigorenko, K. S. Novoselov, T.
J. Booth, T. Stauber, N. M. R. Peres and A. K. Geim, Science, \textbf{320},
1308 (2008).

\bibitem{balandin} A. A. Balandin, S. Ghosh, W. Bao, I. Calizo, D.
Teweldebrhan, F. Miao, C. N. Lau, Nano Lett. \textbf{8}, 902 907 (2008).

\bibitem{ghosh} S. Ghosh, W. Bao, D. L. Nika, S. Subrina, E. P. Pokatilov,
C. N. Lau and A. A. Balandin, Nature Material, \textbf{9,} 555 (2010).

\bibitem{tersoff} D. W. Brenner, Phys. Rev. B \textbf{42}, 9458 (1990).

\bibitem{girifalco} L. A. Girifalco, M. Hodak, R. S. Lee, Phys Rev B,
\textbf{62}, 13104 (2000).

\bibitem{zhong} W. R. Zhong, M. P. Zhang, B. Q. Ai and D. Q. Zheng, Appl.
Phys. Lett., \textbf{98} 113107 (2011).

\bibitem{rafii} H. Rafii-Tabar, Computational Physics of Carbon Nanotubes,
Cambridge University Press, New York (2008).
\end{thebibliography}
\end{document}